\documentclass[epj]{svjour}
\usepackage{psfig}
\begin{document}
\title{Coherent $\omega$ photoproduction  from nuclei and
$\omega$ properties in nuclear matter.}

\author{A. Sibirtsev\inst{1}, Ch. Elster\inst{1,2}  
\and J. Speth\inst{1}}
\institute{Institut f\"ur Kernphysik, Forschungszentrum J\"ulich,
D-52425 J\"ulich \and
Institute of Nuclear and Particle Physics,
Ohio University, Athens, OH 45701}
\date{Received: date / Revised version: date}

\abstract{
The coherent $\omega$-meson photoproduction from nuclei is 
proposed as a phenomenological  method to evaluate the  momentum 
dependence of $\omega$-meson mass shift and width in nuclear matter. 
We analyze available data on un-separated coherent and incoherent 
$\omega$-meson photoproduction from nuclei and extract the imaginary part 
of the  the complex forward $\omega{N}$ scattering amplitude, which
is proportional to the in-medium $\omega$-meson width. 
The accuracy of the currently available data is not sufficient to 
evaluate the real part of forward $\omega{N}$ scattering amplitude 
and reconstruct the momentum dependence of the mass shift of the
$\omega$-meson.
}

\maketitle

\section{Introduction}
The properties of $\rho$ and $\omega$ mesons in a nuclear medium 
remain a topic of extreme interest since the first experimental 
observations by the CERES and HELIOUS-3 Collaborations report
of a strong enhancement of  low mass dilepton production from heavy ion 
collisions~\cite{Agakichiev1,Agakichiev2,Agakichiev3,Masera}.   
This enhancement was interpreted as a modification of the
mass and width of the $\rho$ and $\omega$ mesons in dense 
hadronic matter~\cite{Brown,Li,Rapp,Cassing}. The 
$\rho$-meson is quite broad in the vacuum and an additional increase
of its width in the hadronic environment leads to the production
of an almost continuum dilepton spectrum, which  does not indicate a
resonance structure  and can be identified as perpetual spectral 
strength of the isovector current. 

While the $\rho$-meson seems to be absolutely melted in matter, 
one  expects that the signal of the $\omega$-meson  might survive as a 
quasi resonance structure, even if being strongly modified.
More dedicated experiments aim to isolate the $\omega$-meson
produced in the nuclear medium either by high resolution dilepton 
spectroscopy~\cite{Friese1,Friese2,Stroth,Morrison,Zajc} or through 
the $\omega{\to}\pi^0\gamma$ decay mode~\cite{Messchendorp,Sibirtsev4}.
The goal of these experiments is to identify the structure
of the $\omega$-meson  and to measure  its width and pole position.

The experimental studies of the in-medium modification of the $\omega$-meson 
mass and width are planned by utilizing pion, proton and heavy ion beams 
with HADES at GSI~\cite{Friese1,Friese2,Stroth,Metag1,Bratkovskaya3}, with 
photo-nuclear reactions~\cite{Messchendorp,Effenberger} at 
TAPS~\cite{Gabler} and Crystal Barrel~\cite{Aker} at ELSA,
by heavy ion collisions with PHENIX at RHIC~\cite{Morrison,Zajc} and with 
proton-nucleus reactions by ANKE at COSY~\cite{Sibirtsev4,Golubeva}.

Very recently the E325 experiment at KEK-PS reported~\cite{Ozawa}
new results on the $\rho{/}\omega$ meson modification in
nuclear matter. It was found that the dielectron spectra  produced 
in $p{+}C$ and $p{+}Cu$  collisions at proton beam energy of 12~GeV indicate 
a significant difference  below the $\omega$-meson mass.
The enhancement of the invariant mass spectra of electron-positron pairs
produced from a heavy $Cu$ target might suggest the modification of the
$\rho{/}\omega$ properties at normal nuclear density and for 
average $\omega$-meson momenta  $k_\omega{\simeq}$1~GeV. 

Obviously, apart of the heavy ion reactions, in the above mentioned  
experiments the $\omega$-mesons are  produced with a certain laboratory 
momenta $k_\omega$, that can be quite large. The scale of $k_\omega$ 
is defined by the threshold of the relevant elementary $\omega$-meson 
production reaction. On the other hand, currently available 
predictions~\cite{Tsushima,Friman1,Friman2,Klingl1,Klingl2,Saito}
for the in-medium modification of the $\omega$-meson mass and width 
are given for the $\omega$-meson at rest, i.e. for
momentum $k_\omega{=}0$.

Only some estimates are presently  available~\cite{Friman1,Friman2,Lykasov} 
for finite $k_\omega$ and they are model dependent. 
These predictions~\cite{Friman1,Friman2,Lykasov} are obtained by 
calculations of a forward $\omega{N}$ scattering amplitude
$f_{\omega{N}}$, which within the low density approximation 
is related to an additional in-medium collisional $\omega$-meson
width $\Delta\Gamma_\omega$ 
given by~\cite{Lenz,Dover,Friman1,Friman2}
\begin{equation}
\Delta\Gamma_\omega =4\pi \,\rho_B  \frac{m_n+m_\omega}{m_n m_\omega}
\, \Im f_{\omega{N}}(0),
\end{equation}
where $m_N$ and $m_\omega$ are the free nucleon and $\omega$-meson
masses, respectively, while $\rho_B$ is the baryon density.
At the same time the  $\omega$-meson mass shift $\Delta{m}_\omega$
in nuclear matter is given by~\cite{Lenz,Dover,Klingl2}
\begin{equation}
\Delta{m}_\omega = -2\pi\, \rho_B\, \frac{m_n+m_\omega}{m_n m_\omega}
\, \Re f_{\omega{N}}(0).
\end{equation}

Here we propose phenomenological method to evaluate the forward 
$\omega{N}$ scattering amplitude by coherent $\omega$-meson 
photoproduction off nuclei and to reconstruct the momentum dependence of 
the $\omega$-meson mass and width in nuclear matter. Our paper 
is organized as follows. In Sect.~2 we provide the formalism of 
the $\omega$-meson coherent photoproduction from nuclei.
The analysis of available data as well as an estimation of the incoherent
background are given in Sect.~3. In Sect.~4 we evaluate the
forward scattering amplitude and compare our results with 
the predictions from available sources. The section ends with 
summary of our results.

\section{Coherent $\omega$ photoproduction}
The total  cross section of $\omega$-meson  photoproduction from nuclei, 
$\gamma{+}{A}{\to}\omega{+}{X}$ is given by the sum of the coherent 
and incoherent processes. The 
differential cross section  $d\sigma_{\gamma{A}}^{coh}{/}dt$  for the 
coherent $\omega$-meson photoproduction from nuclei can 
be written in the eikonal form   
as~\cite{Ross,Drell,Kolbig,Bauer,Trefil1,Trefil2} 
\begin{eqnarray}
\frac{d\sigma_{\gamma A}^{coh}}{dt}= 
\left| 2\pi \, f_{\gamma{N}}^{dif}(0)
\!\int\limits^\infty_0\!db
\, J_0(k_tb)\,\,  b \right. \nonumber \\ \left. \times
\!\!\int\limits^\infty_{-\infty}\!\!dz \,\rho(b,z)
  \exp{[ik_lz{+}i\chi_{\omega}(b)]}
\right|^2,
\label{ampl1}
\end{eqnarray}
where an integration is performed over the impact parameter $b$ and 
$z$ coordinate along the direction of the incident photon, 
$\rho(r{=}\sqrt{b^2{+}z^2})$ 
is the nuclear density function normalized to the total nucleon number 
$A$, while $k_l$,  $k_t$ are the longitudinal and transverse  
component of the momentum transfered to the nucleus, respectively,
given by
\begin{equation}
k_l{=}k_\omega-\sqrt{k_\omega^2-m_\omega^2}, \hspace{5mm}
k_t{=}2k_\omega \sin{(\theta  /2)},
\end{equation}
with $k_\omega$ is the total  momentum, $m_\omega$ is the
pole mass of the $\omega$-meson and $t$ is the squared four momentum 
transfered from the photon to $\omega$-meson. In (\ref{ampl1})
$J_0$ is the zeroth order Bessel function and $\chi_{\omega}$
is the nuclear phase shift related to $\omega$-meson 
distortion in nucleus. Within the $t\rho$-approximation this phase
shift can be  approximated by
\begin{equation}
\chi_{\omega}(b) =  \frac{2\pi \, f_{\omega{N}}(0)}{k_\omega}
\int\limits^\infty_z\!\rho(b,y)\, dy,
\end{equation}
where $f_{\omega{N}}(0)$ is complex amplitude for the forward
$\omega{N}$ elastic scattering. The imaginary part of 
$f_{\omega{N}}(0)$ is related by an optical theorem to the
total cross section $\sigma_{\omega{N}}$ of the $\omega{N}$ 
interaction as
\begin{equation}
\Im f_{\omega{N}}(0) =\frac{k_\omega}{4\pi} \, \sigma_{\omega{N}}.
\label{optic}
\end{equation}
Now we introduce the ratio of the real to imaginary part of the 
forward scattering amplitude $\alpha_\omega$ defined as
\begin{equation}
\alpha_\omega=\frac{\Re f_{\omega{N}}(0)}{\Im f_{\omega{N}}(0)}
\label{ratio}
\end{equation}
and finally express the phase shift  $\chi_{\omega}$  as
\begin{equation}
\chi_{\omega} =\frac{\sigma_{\omega{N}} (i+\alpha_\omega)}{2}
\int\limits^\infty_z\!\rho(b,y)\, dy.
\end{equation}

The quantity $f_{\gamma{N}}^{dif}(0)$ in (\ref{ampl1}) 
is the forward diffractive $\omega$-meson photoproduction 
amplitude  for a single nucleon related to the diffractive
cross section $d\sigma_{\gamma{N}}^{dif}{/}dt$ at $t{=}0$ as
\begin{equation}
\left. \frac{d\sigma_{\gamma N}^{dif}}{dt}\right|_{t=0}
=|\,f_{\gamma{N}}^{dif}(0)\,|^2,
\label{difram}
\end{equation}
ant can be expressed in a vector dominance model VDM in terms of 
the forward $\omega{N}$ scattering amplitude $f_{\omega{N}}(0)$. 

The total invariant amplitude ${\cal M}_{\gamma\omega}$ for the 
 reaction $\gamma{+}{N}{\to}{+}\omega{N}$is given by a VDM as
\begin{equation}
{\cal M}_{\gamma\omega} = [
\sum_{V=\rho,\phi,J/\Psi, ...} \!\!\!\!\!\!\frac{\sqrt{\pi \alpha}}
{\gamma_V} {\cal M}_{V\omega}] \,+
\frac{\sqrt{\pi \alpha}}{\gamma_\omega}{\cal{M}_{\omega\omega}},
\label{vdm1}
\end{equation}
where the summation is performed over the available vector meson 
states $V$. The quantity $\alpha$ represents  the fine structure constant,  
$\gamma_V$ denotes the photon coupling to the vector meson state 
and ${\cal M}_{V\omega}$ is the amplitude for the 
transition $V{+}{N}{\to}\omega{+}{N}$ on a nucleon. Moreover, in (\ref{vdm1})
$\gamma_\omega{=}8.24{\pm}0.24$ is the photon-$\omega$ 
coupling constant evaluated~\cite{Sibirtsev1} from the leptonic decays 
of the $\omega$-meson, and ${\cal{M}_{\omega\omega}}$ is an 
invariant amplitude for the  elastic scattering
$\omega{+}{N}{\to}\omega{+}{N}$. 

In principle, (\ref{vdm1}) might also
contain the sum given by a non-diagonal transition of non-vector meson
states to the $\omega$-meson as well as coupling of these possible
intermediate continuum states to the photon. However, only the second
term of (\ref{vdm1}) provides diffractive $\omega$-meson
photoproduction on a nucleon and enters (\ref{ampl1}).

The differential cross section $\gamma{+}{N}{\to}\omega{+}{N}$ is
given by an invariant amplitude ${\cal M}$ as
\begin{equation}
\frac{d\sigma_{\gamma{N}}}{dt} = \frac{|{\cal M}_{\gamma\omega}|^2}
{64 \pi s q^2_\gamma},
\label{dsdt1}
\end{equation}
where $s$ is the squared invariant collision energy and $q_\gamma$ is 
photon momentum in the center of mass system. Now, the
forward diffractive $\omega$-meson photoproduction cross section 
on a single nucleon,  $d\sigma_{\gamma{N}}{/}dt$ at $t{=}0$   
can be expressed in terms of the $\omega{N}$ forward scattering 
invariant amplitude ${\cal{M}_{\omega\omega}}(0)$  as
\begin{equation}
\left. \frac{d\sigma_{\gamma{N}}^{dif}}{dt}\right|_{t=0} 
= \frac{1}{64 s q^2_\gamma}
\frac{\alpha}{\gamma_\omega^2}
|{\cal M}_{\omega\omega}(0)|^2.
\label{forw1}
\end{equation}

Furthermore, the  scattering amplitude $f_{\omega{N}}$ 
of (\ref{optic}) is related to the invariant scattering 
amplitude ${\cal M}_{\omega\omega}$ as
\begin{equation}
f_{\omega{N}} = \frac{1}{8 \pi \sqrt{s}}\, \frac{k_\omega}{q_\omega} 
{\cal M}_{\omega\omega},
\label{labin}
\end{equation}
where $k_\omega$ and $q_\omega$ are the $\omega$-meson momenta in 
the laboratory and the $\omega{N}$ center of mass system, respectively.

Finally, substituting (\ref{optic}) and (\ref{labin}) into 
(\ref{forw1}) the differential cross section for the forward
$\omega$ meson photoproduction on a nucleon is given by
\begin{equation}
\left. \frac{d\sigma_{\gamma{N}}^{dif}}{dt}\right|_{t=0} =
\frac{\alpha}{16 \gamma_\omega^2} \, \frac{q_\omega^2}{q_\gamma^2}
\, (1+\alpha_\omega^2) \, \sigma_{\omega{N}}^2.
\label{dsdt2}
\end{equation}
The differential cross section $d\sigma_{\gamma A}^{coh}{/}dt$  
for the coherent $\omega$-meson photoproduction from nuclei
can then be written as
\begin{eqnarray}
\frac{d\sigma_{\gamma A}^{coh}}{dt}{=}\frac{\pi \, \alpha  \,
\sigma_{\omega{N}}^2} {4 \gamma_\omega^2} \,
\frac{q_\omega^2}{q_\gamma^2}  \left|(i{+}\alpha_\omega)
\int\limits^\infty_0\!db
\, J_0(k_tb)\,  b \!\!\int\limits^\infty_{-\infty}\!\!dz \,\rho(b,z)
\right. \nonumber \\ \left.\times \exp{[ik_lz]} \,
exp\left[\frac{\sigma_{\omega{N}}\, (i\alpha_\omega{-}1)}{2}
\int\limits^\infty_z\!\rho(b,y)\, dy\right]\right|^2\!\!\!\!.\,\,\,\,\,
\,\,\,\,\,
\label{finalc}
\end{eqnarray}

The parameters $\alpha_\omega$ and $\sigma_{\omega{N}}$ can be 
extracted from experimental data on coherent $\omega$-meson
photoproduction from nuclei and through (\ref{optic}) and (\ref{ratio})
should be converted to the real and imaginary part of the
forward $\omega{N}$ scattering amplitude. Furthermore, the coherent
photoproduction can be measured at various photon beam momenta, thus
providing the evaluation of the momentum dependence of the complex 
forward scattering amplitude $f_{\omega{N}}(0)$. In this way 
the momentum dependent potential of the 
$\omega$-meson in nuclear matter can be reconstructed

Eq.(\ref{finalc}) illustrates that the absolute value of the
$\omega$-meson coherent photoproduction differential cross 
section is very sensitive to the size of the total $\omega{N}$ 
cross section, $\sigma_{\omega{N}}$, and thus to the
imaginary part of the forward scattering amplitude
$\Im f_{\omega{N}}(0)$. Therefore $\Im f_{\omega{N}}(0)$
can be uniquely fixed by measuring  the coherent
$\omega$-meson photoproduction from nuclei.

To investigate the  sensitivity of the coherent
$\omega$-meson photoproduction on the real part of the
forward scattering amplitude we calculate the coherent
$\omega$-meson photoproduction cross section from
$Cu$ nuclei at a photon beam energy of $E_\gamma$=6.8~GeV. In the
calculations we use the nuclear density function $\rho{(r})$ 
given by a Wood-Saxon distribution as
\begin{equation}
\rho (r) =\frac{\rho_0}{1+\exp{[\, (r-R)/d\, ]}},
\label{ws}
\end{equation} 
with the nuclear radius being parameterized as
\begin{equation}
R=1.28A^{1/3}-0.76+0.8A^{-1/3} \, \, \, {\mbox fm},
\end{equation}
and the diffusion parameter given by $d{=}\sqrt{3}{/}3$~fm.
Furthermore, for the moment we fix $\sigma_{\omega{N}}$
at 35~mb, as will be motivated later.

\begin{figure}[t]
\vspace*{-3mm}
\hspace*{-2mm}\psfig{file=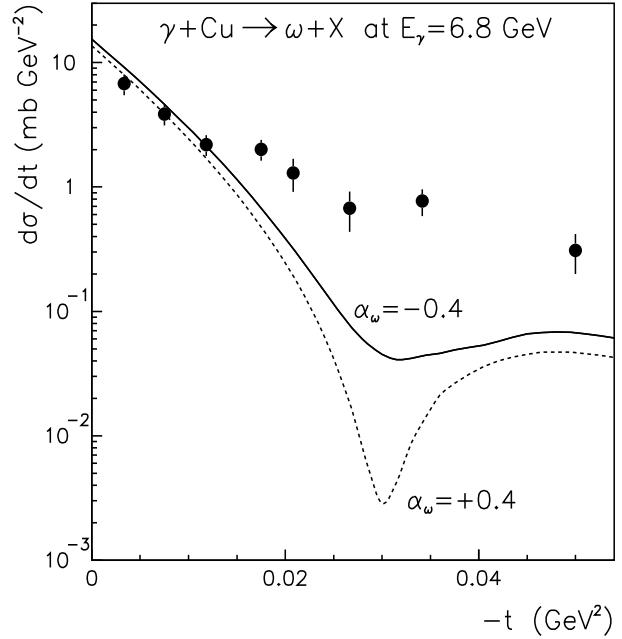,width=9.2cm,height=9.5cm}
\vspace*{-5mm}
\caption{The differential cross section for $\omega$-meson 
photoproduction from $Cu$ target at photon energy  $E_\gamma$=6.8~GeV as 
a function of the squared four momentum transfer $t$. The  lines
show the calculations for the coherent $\omega$-meson photoproduction
by Eq.\protect\ref{finalc} with 
$\sigma_{\omega{N}}$=46~mb and the ratio of the real to imaginary part of
the $\omega{N}$ forward scattering amplitude $\alpha_\omega$=0.4 (dashed)
and $\alpha_\omega$=--0.4 (solid). The solid circles show the  
data~\cite{Behrend} for the $\omega$-meson photoproduction 
where the contributions from both  coherent and incoherent 
processes were not separated experimentally.}
\label{vector2c}
\end{figure}

Fig.\ref{vector2c} shows  ${d\sigma_{\gamma A}^{coh}}{/}{dt}$ 
calculated by (\ref{finalc}) as a function of the 
square of four momentum transfer $t$. The solid line indicates the result 
calculated with the ratio of the real to imaginary part of 
the $f_{\omega{N}}(0)$ taken as $\alpha_\omega$=--0.4, while the
dashed line is the result for  $\alpha_\omega$=0.4. It is clear that 
the ratio $\alpha_\omega$ can be fixed only at the vicinity of the 
diffractive minima. Moreover, at the diffractive minima the 
calculated coherent 
$\omega$-meson photoproduction cross section indicates a substantial
dependence on both, the  absolute value and the sign of the
ratio $\alpha_\omega$. 

\begin{figure}[t]
\vspace*{-3mm}
\hspace*{-1mm}\psfig{file=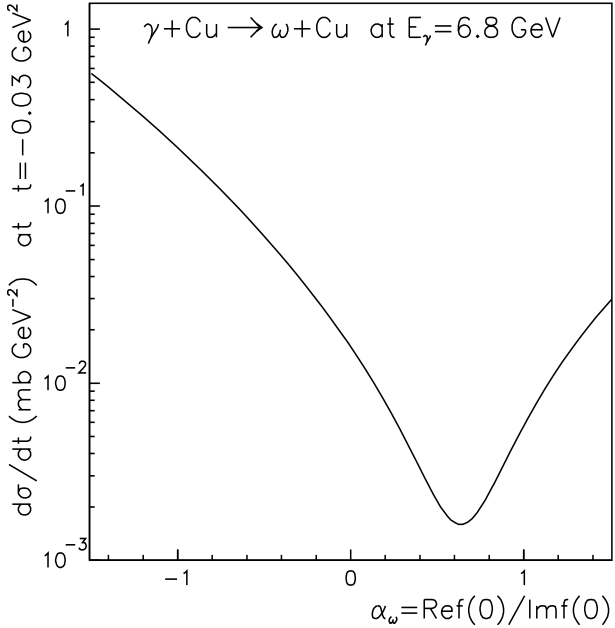,width=9.2cm,height=9.5cm}
\vspace*{-5mm}
\caption{The differential cross section for the coherent $\omega$-meson
photoproduction from $Cu$ target at photon energy $E_\gamma$=6.8 GeV 
calculated around diffractive minima at $t$=--0.03~GeV$^2$ as
a function of the ratio $\alpha$.}
\label{vector2d}
\end{figure}

Furthermore, Fig.\ref{vector2d} shows the differential cross section for 
the coherent photoproduction of $\omega$-mesons from a $Cu$ target calculated
around the diffractive minima at $t$= --0.03~GeV$^2$ as a function of
ratio $\alpha_\omega$. We found that around  the difractive minima
the ${d\sigma_{\gamma Cu}^{coh}}{/}{dt}$ is very sensitive to the 
$\alpha_\omega$ and therefore, the real part of the forward scattering
amplitude $\Re f_{\omega{N}}(0)$ can be directly 
measured by  coherent $\omega$-meson photoproduction. It is 
worthwhile to notice that this method allows a determination of the 
sign of $\Re f_{\omega{N}}(0)$.

The solid circles in Fig.\ref{vector2c} show experimental 
results for $\omega$-meson photoproduction at an average photon 
energy $E_\gamma$=6.8~GeV obtained by Behrend et al.~\cite{Behrend}
at the Cornell electron synchrotron. The produced $\omega$-mesons
were detected through the $\omega{\to}\pi^+\pi^-\pi^0$ decay mode,
thus the measurements~\cite{Behrend} provide the data on inclusive 
$\omega$-meson photoproduction. The contributions from
coherent and incoherent $\omega$-mesons photoproduction were
not separated experimentally~\cite{Behrend}. 

The experimental results on $t$-distribution shown by 
Fig.\ref{vector2c} indicate a forward peak typical for the coherent
$\omega$-meson photoproduction. The photoproduction 
at large $|t|$ is dominated by incoherent production mechanism.
Fig.\ref{vector2c} illustrates that our calculations based on 
(\ref{finalc}) reasonably reproduce the experimental results~\cite{Behrend}
at small $|t|$ and manifistate the dominance of the coherent photoproduction
at $t{\ge}$--0.01~GeV$^2$. The data~\cite{Behrend} can be
used only for the evaluation of $\sigma_{\omega{N}}$, or imaginary
part of the forward $\omega{N}$ scattering amplitude.

\section{Data analysis}
As it  was shown in Fig.~\ref{vector2c} the comparison between
our calculations and the experimental results  requires an 
estimate of the contribution from the incoherent $\omega$-meson 
photoproduction. Moreover, the evaluation of the incoherent
background is important for an determination of the
experimental efficiency necessary for the separation
of the coherent and incoherent $\omega$-meson photoproduction, 
especially near the difractive minima, where the coherent cross section
is sensitive to the real part of the $\Re f_{\omega{N}}(0)$.

Actually, the results shown in Fig.~\ref{vector2c} indicate
that for $Cu$ targets and a photon energy of $E_\gamma$=6.8~GeV 
the contribution from incoherent $\omega$-meson photoproduction
around $t{\simeq}$--0.03~GeV$^2$ may exceed the coherent one
by two or three orders of magnitude. Whether this large difference
holds for other nuclear targets and photon energies 
requires an additional calculation of the incoherent $\omega$-meson
photoproduction. 

Taking into account only the leading term in $t$  given by 
$\omega$-meson photoproduction and the absorption in the nucleus the 
differential cross section ${d\sigma_{\gamma A}^{inc}}{/}{dt}$ 
for incoherent $\omega$-meson photoproduction is given by~\cite{Kolbig}
\begin{eqnarray}
\frac{d\sigma_{\gamma A}^{inc}}{dt}=
\left. \frac{d\sigma_{\gamma N}^{tot}}{dt}\right|_{t=0}\,
\!\!\!\!\!\!\!\!\exp{(at)}\,   \frac{2\pi}{\sigma_{\omega{N}}}
\int\limits^\infty_0  b \,  db \nonumber \\
\times ( 1-\exp{\!\!\left[ -\sigma_{\omega{N}}
\!\!\int\limits^\infty_{-\infty}
\!\!\rho(b,y)dy\right]}),
\label{incoh}
\end{eqnarray}
where $a$ is the slope of $t$-distribution from the elementary
reaction $\gamma{+}{N}{\to}\omega{+}{N}$. Furthermore, the
elementary forward $\omega$-meson photoproduction cross section
${d\sigma_{\gamma N}^{tot}}{/}{dt}$ at $t{=}0$ now should account not 
only for the diffractive  process due to the 
$\omega{+}{N}{\to}\omega{+}{N}$
transition as is given by Eq.\ref{dsdt2}, but also for the total 
$\gamma{N}{\to}\omega{N}$ amplitude of (\ref{vdm1}).

\begin{figure}[b]
\vspace*{-5mm}
\hspace*{-2mm}\psfig{file=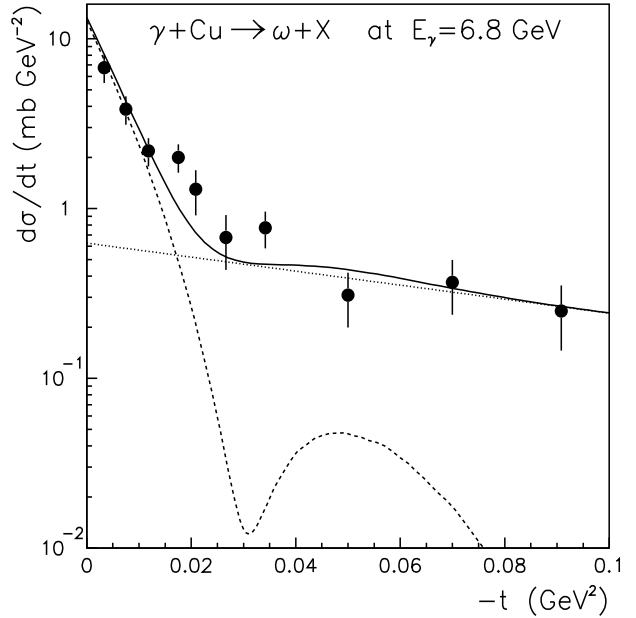,width=9.2cm,height=9.4cm}
\vspace*{-5mm}
\caption{The differential cross section for $\omega$-meson 
photoproduction from $Cu$ nucleus at photon energy  $E_\gamma$=6.8~GeV as 
function of the squared four momentum transfer $t$. The solid circles 
show the  data from Ref.~\protect\cite{Behrend}. The dashed line
indicates the calculations for the coherent $\omega$-meson photoproduction
by Eq.\protect\ref{finalc}, while the dotted line is the incoherent
$\omega$ photoproduction cross section given by Eq.\ref{incoh}.
The solid line shows their sum. The calculations were performed 
with $\sigma_{\omega{N}}$=46~mb, the ratio  $\alpha_\omega$=0
and slope $a$=9.5~GeV$^{-2}$.}
\label{vector2a}
\end{figure}

However, as will be shown later, at high energies 
(\ref{dsdt2}) accounts for the substantial part of a
total forward $\gamma{+}{N}\to\omega{+}{N}$ cross section.
On the other hand, ${d\sigma_{\gamma N}^{tot}}{/}{dt}$ at $t{=}0$ 
as well as slope $a$ can be evaluated from experimental 
data~\cite{LB}. 

Moreover, higher order corrections to (\ref{incoh}) due to the 
single elastic scattering after the $\omega$-photoproduction 
on a target nucleon are proportional to the powers of $\exp{(at)}$ 
and might be important only at large $|t|$. However, this is beyond the 
region of our interest. We also do not consider photon 
shadowing, which might play certain role in the incoherent 
$\omega$-meson photoproduction. 

The differential cross section for
$\gamma{+}{Cu}{\to}\omega{+}{X}$ reaction at $E_\gamma$=6.8~GeV. 
is shown in Fig.~\ref{vector2a}. Solid
circles represent experimental results from Ref.~\cite{Behrend}. The lines 
show the calculations performed with $\sigma_{\omega{N}}$=46~mb, 
the ratio  $\alpha_\omega$=0 and a slope $a$=9.5~GeV$^{-2}$~\cite{LB}.
Here we also take ${d\sigma_{\gamma N}^{tot}}{/}{dt}$ at $t{=}0$ 
as given by (\ref{dsdt2}). 
The dashed line in Fig.~\ref{vector2a} shows our result for
the differential cross section for coherent $\omega$-meson photoproduction
calculated by (\ref{finalc}). The dotted line
indicates the result for the incoherent $\omega$-meson photoproduction as
calculated by (\ref{incoh}). The solid line is the sum of
coherent and incoherent processes and reasonably describes the
data~\cite{Behrend}.

\begin{figure}[t]
\vspace*{-3mm}
\hspace*{-2mm}\psfig{file=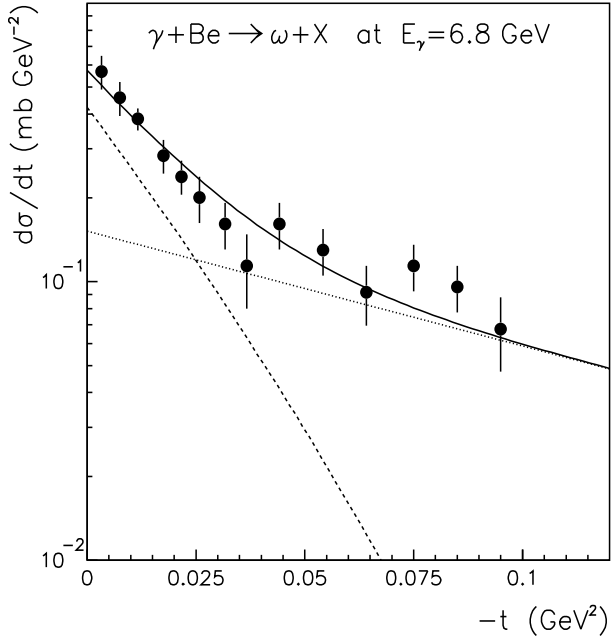,width=9.2cm,height=9.5cm}
\vspace*{-5mm}
\caption{The differential cross section for $\gamma{Be}{\to}\omega{X}$
reaction at $E_\gamma$=6.8~GeV as 
a function of the squared four momentum transfer $t$. The solid circles 
show the  data from Ref.~\protect\cite{Behrend}. The dashed line
indicates the calculations for the coherent $\omega$-meson photoproduction
by Eq.\protect\ref{finalc},  the dotted line is  incoherent
$\omega$ photoproduction cross section given by Eq.\ref{incoh},while
the solid line shows their sum. The calculations were performed 
with $\sigma_{\omega{N}}$=46~mb, the ratio  $\alpha_\omega$=0
and slope $a$=9.5~GeV$^{-2}$.}
\label{vector2b}
\end{figure}

Fig.~\ref{vector2b} shows the differential cross section for 
$\omega$-meson photoproduction from $Be$ at a photon energy
$E_\gamma$=6.8~GeV. Again, the experimental results are taken from
Ref.~\cite{Behrend} and the lines show the calculations
with $\sigma_{\omega{N}}$=46~mb, $\alpha_\omega$=0 and 
$a$=9.5~GeV$^{-2}$~\cite{LB}. The results for coherent and
incoherent $\omega$-meson photoproduction are shown by the dashed 
and dotted lines, respectively. The solid line indicates their
sum and describes the data reasonably well. Note that the position of the
diffractive minima for light nuclei lies at substantially larger
$|t|$~\cite{Sibirtsev2}. 

In the calculations for $Be$  we use the density
function given by~\cite{Dalkarov}
\begin{equation}
\rho (r) = \frac{1}{(R\sqrt{\pi})^3}\left[4+
\frac{2(A-4)r^2}{3R^2}\right]\exp{\left[-\frac{r^2}{R^2}\right]},
\end{equation}
where $R$=1.58~fm.

The $\omega$-meson photoproduction from nuclei at
average photon energies $E_\gamma$=8.2~GeV was studied experimentally 
by Aramson et al.~\cite{Abramson} with  results compatible
to their previous measurements given at $E_\gamma$=6.8~GeV
in Ref.~\cite{Behrend}.

In addition, the $\omega$-meson nuclear photoproduction at 
average photon energies $E_\gamma$=5.7~GeV was measured at 
DESY~\cite{Braccini}  by detecting the 
$\omega{\to}\pi^0\gamma$ decay. The coherent and incoherent
processes were not separated and the data were published without
absolute normalization. 

\begin{figure}[b]
\vspace*{-9mm}
\hspace*{-2mm}\psfig{file=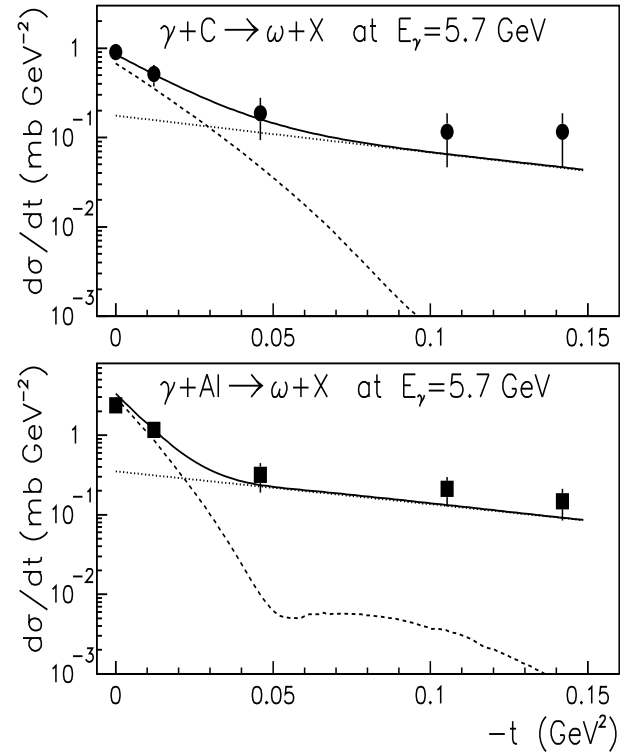,width=9.2cm,height=11.4cm}
\vspace*{-7mm}
\caption{The differential cross section for $\omega$-meson 
photoproduction from $C$ and $Al$ at $E_\gamma$=5.7~GeV as a function of 
the squared four momentum transfer $t$. The data were 
taken from Ref.~\protect\cite{Braccini}. 
The lines show our results for the coherent (dashed), incoherent (dotted)
and the sum (solid) of coherent and incoherent photoproduction
given by Eqs.\protect\ref{finalc},\ref{incoh} and
calculated with  $\sigma_{\omega{N}}$=46~mb, ratio $\alpha_\omega$=0
and slope $a$=9.5~GeV$^{-2}$.}
\label{vector4}
\end{figure}

Fig.\ref{vector4} shows the differential cross section for
$\omega$-meson photoproduction from $C$ and $Al$ targets at 
photon beam energy $E_\gamma$=5.7~GeV. The DESY experimental results are 
taken from Ref.~\cite{Braccini} and normalized by our calculations.
The solid lines show the 
sum of the coherent and incoherent photoproduction cross sections.
The dotted lines show the calculations for incoherent $\omega$-meson
photoproduction, which is dominant at large $|t|$. The contribution from
coherent $\omega$-meson photoproduction is given by the dashed lines.
The calculations were performed using (\ref{finalc}) and (\ref{incoh})
with parameters  $\sigma_{\omega{N}}$=46~mb, $\alpha_\omega$=0
and $a$=9.5~GeV$^{-2}$~\cite{LB}. 

The data~\cite{Braccini} are  quite reasonably reproduced by the 
calculations. However, they are  insensitive to the ratio 
$\alpha$. Our choice of $\alpha_\omega$=0 for the comparison
between the calculations and the data~\cite{Behrend,Braccini}
is indeed arbitrary. We conclude that the ratio $\alpha$ as well
as the real part of the forward $\omega{N}$ scattering amplitude
$\Re f_{\omega{N}}(0)$ can not be constrained by the available
data~\cite{Behrend,Abramson,Braccini}, since these measurements
did not isolate the coherent photoproduction. 

The total coherent and incoherent $\omega$-meson photoproduction
cross sections from complex nuclei at $E_\gamma$=3.9~GeV were measured 
by Brodbeck et al.~\cite{Brodbeck} and extrapolated to $t$=0.
Experimental results for $\omega$-meson photoproduction
from nuclei at $E_\gamma$=6.8~GeV and at $t$=0 were also
reported by Behrend et al.~\cite{Behrend1}.

\begin{figure}[t]
\vspace*{-3mm}
\hspace*{-2mm}\psfig{file=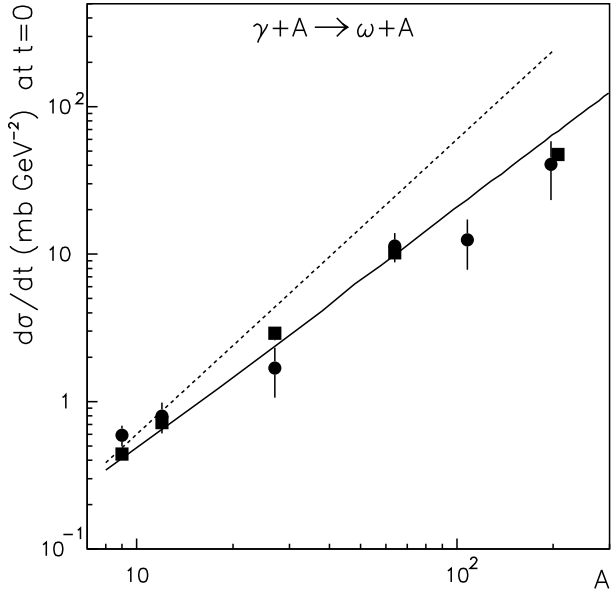,width=9.2cm,height=9.3cm}
\vspace*{-7mm}
\caption{The differential cross section for $\omega$-meson 
photoproduction at $t$=0 as a function of target mass number $A$.
The circles show the data~\cite{Brodbeck} for photon beam energy 
$E_\gamma$=3.9~GeV, while the squares show the 
results~\cite{Behrend1} measured at  $E_\gamma$=6.8~GeV. 
The solid line shows our calculations  with $\sigma_{\omega{N}}$=37~mb
and $\alpha_\omega$=0. The dashed line indicates the $A^2$-dependence
obtained with $\sigma_{\omega{N}}$=0.}
\vspace*{-1mm}
\label{vector5}
\end{figure}

Fig.\ref{vector5} shows   results for $\omega$-meson photoproduction cross 
section at $t$=0 as a function of the target mass number. 
The solid circles indicate data~\cite{Brodbeck} collected at 
$E_\gamma$=3.9~GeV, while the squares show the results~\cite{Behrend1} 
measured at photon energy $E_\gamma$=6.8~GeV. The data do not 
show any dependence on the photon beam energy. 

The solid line in Fig.\ref{vector5} 
indicates our calculations with $\sigma_{\omega{N}}$=37~mb
and ratio $\alpha_\omega$=0. We note that our results do not depend
on the ratio $\alpha$. The dashed line in Fig.~\ref{vector5}
shows the $A^2$-dependence which is given by (\ref{finalc})
assuming that $\sigma_{\omega{N}}$=0 and neglecting the 
$k_l$-dependence. The dashed line is arbitrary normalized 
at $A{=}Be$. 

Moreover, the difference between the solid and dashed lines in
the Fig.~\ref{vector5} is due to the finite value of $\sigma_{\omega{N}}$. 
Therefore, the data~\cite{Brodbeck,Behrend1,Behrend1} 
on forward $\omega$-meson photoproduction provide 
a reasonable evaluation of the imaginary part of the 
$\omega{N}$ forward scattering amplitude $\Im f_{\omega{N}}(0)$ .  

\section{Evaluation of the forward scattering amplitude}
The comparison between our calculations and  experimental
results~\cite{Behrend,Abramson,Braccini,Brodbeck} allows
us to extract the imaginary part of the forward $\omega{N}$ 
scattering. Finally the data on $\omega$-meson 
photoproduction from nuclei were fitted in order to evaluate
$\sigma_{\omega{N}}$. Although we found that the available data
are absolutely insensitive to the ratio $\alpha_\omega$ we allow
its variation by a minimization procedure~\cite{Minuit}.

The results of the minimization are shown in the Table.~\ref{tab1}.
None of the data sets  allows to extract a specific value
for the ratio $\alpha_\omega$. Some of the experimental results
allow an extraction of $\sigma_{\omega{N}}$ with sufficiently 
large uncertainty.  

However, we should  clarify, that experimental
results are insensitive to the ratio $\alpha_\omega$ in case 
it is less than one. As is shown by (\ref{finalc}) the large values
of $\alpha_\omega$ should immediately effect our results.
Roughly, a large ratio $\alpha_\omega$ should cause
a decrease of the total $\omega{N}$ cross section $\sigma_{\omega{N}}$.
The fitting procedure does not indicate such a tendency. Our
analysis suggest that $\alpha_\omega{<}1$ or 
$\Im f_{\omega{N}}(0){>}$$\Re f_{\omega{N}}(0)$.
 
Moreover, the $\omega$-meson photoproduction data on $A$-dependence
at $t$=0 shown in the Fig.\ref{vector5} show substantial sensitivity 
to $\sigma_{\omega{N}}$.
Obviously, the situation might improve substantially when  separate 
data for coherent photoproduction will be available.

\begin{table}[h]
\caption{The total $\omega{N}$ cross section $\sigma_{\omega{N}}$
in mb and the ratio of the real to imaginary $\omega{N}$ scattering 
amplitude $\alpha_\omega$ evaluated from the data on $\omega$-meson 
photoproduction from nuclei at photon energy $E_\gamma$
given in GeV. Also are shown the total $\chi^2$, number 
of experimental points $nep$ and the legend and the reference 
to experimental results.}
\label{tab2}
\begin{tabular}{ccccccc}
\hline\noalign{\smallskip}
$E_\gamma$  & $\sigma_{\omega{N}}$ & $\alpha_\omega$
& $\chi^2$ & $nep$ & \,\, legend\,\, & Ref. \\
\noalign{\smallskip}\hline\noalign{\smallskip}
3.9 & 36.1$\pm$2.2 & 0.17$\pm$0.20 & 10.2 & 6 & $t$=0 &
\protect\cite{Brodbeck} \\
6.8 & 36.8$\pm$2.0 & --0.01$\pm$0.03 & 33.8 & 5 & $t$=0 &
\protect\cite{Behrend1}\\
6.8 & 45.9$\pm$1.3 & 0.2$\pm$0.1 & 11.5 & 14 & $Be$& \protect\cite{Behrend}\\
6.8 & 44.8$\pm$0.1 & --0.11$\pm$0.02 & 14.6 & 14 &$Cu$& 
\protect\cite{Behrend}\\
5.7 & 48$\pm$4 & 0.15$\pm$0.18 & 1.4 & 5 &$C$& \protect\cite{Braccini}\\
5.7 & 44$\pm$2 & 0.05$\pm$0.02 & 5.4 & 5 &$Al$& \protect\cite{Braccini}\\
\noalign{\smallskip}\hline
\end{tabular}
\label{tab1}
\end{table}

\begin{figure}[b]
\vspace*{-6mm}
\hspace*{-2mm}\psfig{file=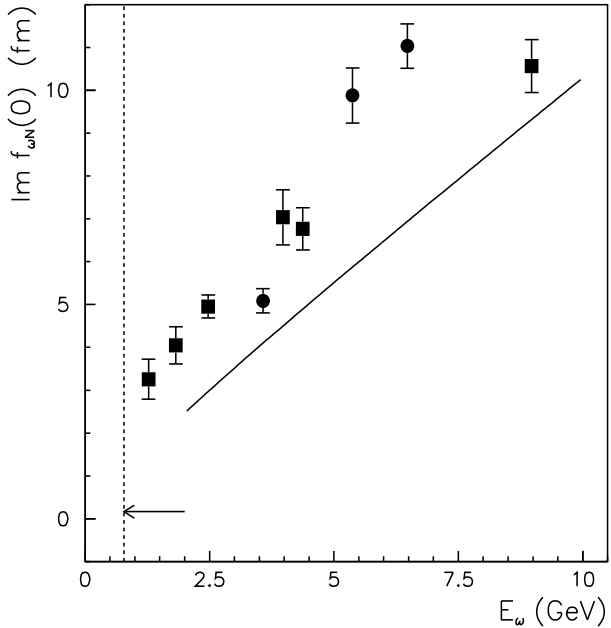,width=9.2cm,height=9.5cm}
\vspace*{-5mm}
\caption{The imaginary part of the forward $\omega{N}$ scattering
amplitude $\Im f_{\omega{N}}(0)$ as a function of  laboratory 
$\omega$-meson total energy $E_\omega$. The solid circles show our results 
extracted from $\omega$-meson photoproduction off nuclei. The squares 
show the result evaluated by (\ref{optic}) and (\ref{dsdt2}) from the 
data~\cite{LB} on forward $\omega$-meson photoproduction 
in $\gamma{+}{p}{\to}\omega{+}{p}$ reaction.
The dashed line shows the $\omega{N}$ threshold given by
$\omega$-meson pole mass. The arrow indicates the lower limit 
for $\Im f_{\omega{N}}(0)$ at threshold obtained~\cite{Hanhart} from 
the $\pi{+}{N}{\to}\omega{+}{N}$ data~\cite{Karami}. The solid line is the 
predictions from additive quark model and Regge theory 
given by (\ref{aqm}).}
\label{vector6}
\end{figure}

Final results for $\sigma_{\omega{N}}$
are converted by (\ref{optic}) to imaginary part of 
$f_{\omega{N}}(0)$ and are shown by the solid circles in 
Fig.\ref{vector6} as a function of the  total laboratory 
$\omega$-meson energy $E_\omega$. In the case, when the results
for $\sigma_{\omega{N}}$ are evaluated from different data sets 
but at the same photon energy, we average them in calculating the
imaginary part of the forward scattering amplitude.

Now our results can be compared with  estimations given by
(\ref{optic}) and (\ref{dsdt2}) and available  data on  forward 
$\omega$-meson photoproduction on a free nucleon.  
The squares in Fig.\ref{vector6} show the $\Im f_{\omega{N}}(0)$
obtained from the data~\cite{LB} on $\gamma{+}{p}{\to}\omega{+}{p}$ 
reaction cross section at $t$=0 and neglecting the ratio 
$\alpha_\omega$ in (\ref{dsdt2}). 

An arrow at Fig.~\ref{vector6} indicates the lower
limit for the imaginary part of the $\omega{N}$ scattering amplitude
at threshold $E_\omega{=}m_\omega$ given at the pole mass of 
$\omega$-meson by
\begin{equation}
\Im f_{\omega{N}}(0) =0.164 \ {\mbox fm}, 
\end{equation}
as recently evaluated~\cite{Hanhart} from  $\pi^-{+}p\to\omega{+}{n}$
measurements~\cite{Karami}.

Moreover, within an additive quark model the forward $\omega{N}$ 
scattering amplitude is given as the average of the forward $\pi^-N$ and 
$\pi^+N$ amplitudes~\cite{Lipkin,Kajantie,Donnachie1}. Using the 
Donnachie and Landshoff results~\cite{Donnachie2} for a Regge 
theory fit to the $\pi{N}$ total cross section the
complex forward $\omega{N}$ scattering amplitude can be 
written as 
\begin{eqnarray}
f_{\omega N}(0)= \frac{k_\omega}{4\pi \, {\hbar}c} \,
[\,  (\, 0.173 \left[ \frac{s}{s_0}\right]^\epsilon{-}
2.726\left[\frac{s}{s_0}\right]^{-\eta})  \nonumber \\ 
+ i
(\, 1.359\left[\frac{s}{s_0}\right]^\epsilon{+}3.164
\left[\frac{s}{s_0}\right]^{-\eta} )\, ]\,\,\, \mbox{fm},
\label{aqm}
\end{eqnarray}  
where the $\omega$-meson momentum $k_\omega$ and squared invariant
collision energy $s$ are given in GeV/c and $s_0$=1~GeV$^2$. The 
effective powers $\epsilon$=0.08 and $\eta$=0.45 are given by
the pomeron and $\rho$, $\omega$, $f$, $a$ exchanges, respectively. 
The solid line in Fig.~\ref{vector6} shows the prediction by additive
quark model calculated by (\ref{aqm}). We find that the prediction from
the additive quark model and the Regge theory systematically
underestimates other results.

Although we do not consider the results for ratio $\alpha_\omega$
listed in the Table~\ref{tab1} at some confidence level, it seems 
worthwhile to discuss a possible estimate for the ratio of the
real to imaginary part of the forward $\omega{N}$ scattering 
amplitude. Moreover, as is illustrated by Fig.\ref{vector2d} the 
magnitude of the coherent differential $\omega$-meson photoproduction 
cross section at diffractive minima substantially depends on the 
both sign and size of the $\alpha_\omega$ and thus the estimate for 
this ratio may be useful for valuation of an experimental 
accuracy necessary for measurements around the diffractive minima.

\begin{figure}[t]
\vspace*{-3mm}
\hspace*{-2mm}\psfig{file=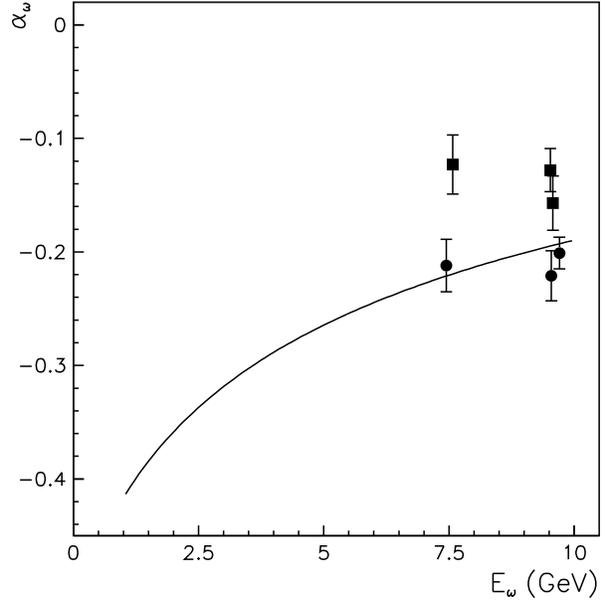,width=9.2cm,height=9.1cm}
\vspace*{-5mm}
\caption{The ratio $\alpha_\omega$ of the real to imaginary part of 
the forward $\omega{N}$ scattering amplitude as a function of total
laboratory $\omega$-meson energy $E_\omega$. The solid 
circles show experimental results~\cite{Foley} for $\pi^+p$ scattering
taken at the same invariant collision energy $\sqrt{s}$, 
while the squares are that for $\pi^-p$ scattering~\cite{Foley}. 
The solid line show the predictions for $\omega{N}$ scattering given 
by an additive quark model and Regge theory~\cite{Donnachie1} and 
calculated by (\ref{aqm}).}
\label{vector7}
\end{figure}

Very rough estimates of the ratio $\alpha_\omega$ may be given
by an additive quark model through (\ref{aqm}) and is shown in
Fig.\ref{vector7} by the solid line as a function of total $\omega$-meson
laboratory energy $E_\omega$. The solid circles in the 
Fig.\ref{vector7} show the experimental results~\cite{Foley} for 
the ratio of the real to imaginary part for the $\pi^-p$ forward 
scattering amplitude while the squares indicate the data for
$\pi^+p$ scattering. The $\pi{N}$ data were taken at the same 
invariant collision energies $\sqrt{s}$.

Furthermore, the results from the coupled channel 
analysis~\cite{Friman1} of the meson-nucleon scattering 
data predicts at the $\omega{N}$ threshold
\begin{equation}
f_{\omega N} =-0.5 +i\, 0.2 \, \,\, \mbox{fm},
\label{frim}
\end{equation}
providing the negative ratio $\alpha_\omega$=--2.5. Moreover, most 
recently within the same model approach~\cite{Friman2} it was deduced 
that $\Im f_{\omega{N}}$=0.013~fm, while the real part 
$\Re f_{\omega{N}}$ stands the same as in Eq.\ref{frim} resulting
finally in the ratio $\alpha_\omega$=--38.5.

At the same time, the calculations~\cite{Klingl1} based on an 
effective Lagrangian approach combined with chiral SU(3) dynamics 
predicts the forward $\omega{N}$ scattering amplitude as
\begin{equation}
f_{\omega N} =3.34 +i\, 2.1 \, \, \, \mbox{fm},
\end{equation}
with a positive defined ratio $\alpha_\omega$=0.62. 
More recent results of the calculation~\cite{Klingl2}   
of the scattering amplitude is given as
\begin{equation}
f_{\omega N} =1.6 +i\, 0.3 \, \,\, \mbox{fm},
\end{equation}
with positive and large ratio $\alpha_\omega$=5.3.
 
Theoretical predictions available for $\alpha_\omega$
differ substantially both in sigh and the absolute
value of the ratio. 
Obviously, the coherent $\omega$-meson photoproduction from
nuclei can not directly resolve the existing discrepancies between 
available predictions for the forward $\omega{N}$ scattering amplitude
exactly at $E_\omega{=}m_\omega$. However the  
$\gamma{+}{A}{\to}\omega{+}{A}$ reactions might be considered as a model 
independent measurements of $f_{\omega{N}}$ at final $\omega$-meson
momenta as well as provide an orientation for analytical continuation 
of the forward scattering amplitude close to the 
threshold~\cite{Eletsky,Kondratyuk,Sibirtsev3}. 

It is important to note that both, the coherent and incoherent 
cross sections for forward $\omega$-meson photoproduction from nuclei
will be influenced when large values of $\alpha{>}$1 are used.

\section{Conclusions}
The coherent $\omega$-meson photoproduction from nuclei is studied 
as  a possible phenomenological method to evaluate the complex forward
$\omega{N}$ scattering amplitude at finite $\omega$-meson
momenta. 

We found that the real part of the forward scattering
amplitude $\Re f_{\omega{N}}(0)$ can be only fixed by the data
on coherent $\omega$-meson photoproduction around the diffractive 
minima. At the same time, the imaginary part of the forward
scattering amplitude $\Im f_{\omega{N}}(0)$ can be well evaluated by 
an absolute value of the cross section at small $|t|$, where the
contribution from incoherent processes is relatively small. Moreover,
the sign and the magnitude of the ratio $\alpha_\omega$ of the real 
to imaginary part of the forward $\omega{N}$ scattering amplitude can 
be  identified by the magnitude of the coherent photoproduction
cross section at diffractive minima. 

We analyze available experimental 
data~\cite{Behrend,Abramson,Braccini,Brodbeck,Behrend1} on $\omega$-meson 
photoproduction from nuclei.  Unfortunately the contributions 
from coherent and incoherent processes are not separated 
experimentally. We find reasonable agreement between the 
data~\cite{Behrend,Abramson,Braccini,Brodbeck,Behrend1} and our 
calculations including both, coherent and incoherent $\omega$-meson
photoproduction. By fitting the experimental results we evaluate 
the imaginary part of the forward $\omega{N}$ scattering amplitude.
Within our approach we do not variate the $\omega$-photon coupling
constant, but fix it by the dileptonic decay. We find that 
data, which are not separated into coherent and incoherent
production, can not provide  reliable results for the 
real part of the scattering amplitude $\Re f_{\omega{N}}(0)$.

Finally, our results for $\Im f_{\omega{N}}(0)$ are compared 
with predictions given by a vector dominance model relation between 
the forward $\gamma{+}{N}{\to}\omega{+}{N}$ reaction cross section and
imaginary part of the forward $\omega{N}$ scattering amplitude.
The experimental data~\cite{LB} for forward $\omega$-meson
photoproduction from the proton provide an $\Im f_{\omega{N}}(0)$
that agrees  reasonably well with our estimates. 

Furthermore, the predictions by an additive quark model together 
with a Regge model fit to the data for hadronic cross sections 
systematically underestimate our results for $\Im f_{\omega{N}}(0)$
as well as that extracted from the $\gamma{+}{p}{\to}\omega{+}{p}$ reaction. 
This discrepancy might indeed indicate that, while  experimental 
results on exclusive $\rho$, $\phi$  and $J{/}\Psi$ photoproduction 
excellently confirm~\cite{Donnachie1,Donnachie2,Sibirtsev3} the
Regge theory, the $\omega$-meson photoproduction could be 
considered as an exception. It is important  that 
$\Im f_{\omega{N}}(0)$ is evaluated precisely from available
data on the $A$-dependence of $\omega$-meson photoproduction at $t$=0.

We also discuss the current status of the predictions for the
real part of the $\omega{N}$ scattering amplitude and detect
a strong disagreement between the results from different models.

\begin{acknowledgement}
This work was performed in part under the auspices of the 
U.~S. Department of Energy under contract No. DE-FG02-93ER40756 
with the Ohio University.
\end{acknowledgement}

\end{document}